\documentclass[pra,twocolumn,superscriptaddress,showpacs]{revtex4-1}
\usepackage{amssymb,amsmath,amsfonts,bm}
\usepackage{epsfig,graphicx}
\usepackage{hyperref}
\usepackage{color}
\usepackage{times}

\begin{document}

\title{Fast holonomic quantum computation based on solid-state spins
 with all-optical control}

\author{Jian Zhou}

\affiliation{Department of Electronic Communication Engineering, Anhui Xinhua University, Hefei, 230088, China}
\affiliation{Guangdong Provincial Key Laboratory of Quantum Engineering and Quantum Materials, and School of Physics\\ and Telecommunication Engineering, South China Normal University, Guangzhou 510006, China}
%\affiliation{National Laboratory of Solid State Microstructure, Nanjing University, Nanjing 210093, China}

\author{Bao-Jie Liu}
\author{Zhuo-Ping Hong}
\author{Zheng-Yuan Xue} \email{Corresponding author. Email: zyxue@scnu.edu.cn}
\affiliation{Guangdong Provincial Key Laboratory of Quantum Engineering and Quantum Materials, and School of Physics\\ and Telecommunication Engineering, South China Normal University, Guangzhou 510006, China}

\date{\today}

\begin{abstract}
Holonomic quantum computation is a quantum computation strategy that promises some built-in noise-resilience features. Here, we propose a scheme for nonadiabatic holonomic quantum computation with nitrogen-vacancy center electron spins, which are characterized by fast quantum gates and long qubit coherence times. By varying the detuning, amplitudes, and phase difference of lasers applied to a nitrogen-vacancy center, one can directly realize an arbitrary single-qubit holonomic gate on the spin. Meanwhile, with the help of cavity-assisted interactions, a nontrivial two-qubit holonomic quantum gate can also be induced. The distinct merit of this scheme is that all the quantum gates are obtained via an all-optical geometric manipulation of the solid-state spins. Therefore, our scheme opens the possibility for robust quantum computation using solid-state spins in an all-optical way.
\end{abstract}

\maketitle

\section{Introduction}
Holonomic quantum computation \cite{Zanardi1999}, in which quantum gates are realized by nonabelian geometric phases, has emerged as a promising way to implement quantum computation. This is because geometric phases depend on certain global properties of the travel path of the Hamiltonian and are thus robust against local fluctuations \cite{dp,Johansson2012, Solinas2012, Jing2017}. However, previous holonomic quantum computation schemes \cite{Duan2001a, Recati2002, ps, Zhang2005c, Zhang2006} usually adopt four-level tripod quantum systems in which the coherent joint control of the time-dependent parameters is needed for the system Hamiltonian; thus, their experimental demonstration is difficult. Moreover, holonomic quantum gates are obtained by adiabatic evolution, which is not preferable, because the time needed may be of the same order as the qubit coherent times for typical quantum systems \cite{Wang2001,Zhu2002}. Therefore, geometric phases via nonadiabatic evolution, in which the adiabatic condition is not required, can support the fast implementation of quantum gates while reserving the noise-resilience features. We also note that another possible solution to the long running time difficulty are the so-called superadiabatic transitionless driving \cite{Berry,TQD1,TQD2} or shortcut to adiabaticity \cite{chenxi,PRL2012} protocols, wherein the adiabatic process is speeded up while still retains its merits. However, there is a need for case studies on different tasks for complicated modification of the driven fields.

Based on fast nonabelian geometric phases, nonadiabatic holonomic quantum computation (NHQC) with three-level systems driven by short resonant laser pulses has been proposed \cite{Wu2005, Sjoqvist2012, Xu2012, vam2014, Cadez2014, Liang2014a, Zhang2014d, Xu2014, Xu2015, Zhou2015, Xue2015b, Xue2016, Zhao2016, Herterich2016,xu2017, vam2017, Xue2017,zhao2017, xu20172, vam2016}. This type of NHQC \cite{Sjoqvist2012} allows for potentially easy experimental implementation of the holonomic quantum gates; thus, it provides  a practical way for implementing this noise-resilient quantum computation. Meanwhile, NHQC has been experimentally demonstrated with superconducting circuits \cite{Abdumalikov2013}, NMR \cite{Feng2013,s1} and nitrogen-vacancy (NV) center electron spins in diamond \cite{Zu2014, Arroyo-Camejo2014,s0,s2}.

Recently, owing to its long electronic spin lifetime, fast initialization and optical readout, and coherent manipulation even at room temperature, the NV center is considered a promising candidate for quantum computing \cite{Togan2010, Shi2010}. However, we note that previous implementations of universal NHQC with NV centers usually employ microwave control of the qubit states and that individual addressing without crosstalk is difficult. Meanwhile, the quantum-gate strategy with microwave control is not compatible to the initialization and readout of the qubits, which are usually achieved optically.

Here, we propose a scheme for universal NHQC with NV center electron spins via all-optical manipulation, where the initialization, readout, and quantum gates of the solid-state spins can be realized using coherent population trapping and stimulated Raman techniques \cite{Yale2013}. Quantum gates acting in a single spin state can be obtained with very high efficiency due to the sufficiently long electronic spin lifetime. By separately varying the detuning and amplitude of the lasers, a set of universal single-qubit gates can be realized by all-optical manipulation on the NV centers \cite{Yale2013, Yale2016,optical1,optical2}. Moreover, with the help of the lowest whispering-gallery mode in an fused-silica microsphere optical cavity or an optical fiber linked cavity structure, a nontrivial two-qubit holonomic quantum gate can be induced between the two involved spins. In addition, the performance of the gates is evaluated by numerical simulation under decoherence. Therefore, our scheme provides a promising alternative for robust quantum computation based on solid-state spins in an all-optical way.

\begin{figure*}[tbp]
\centering
\includegraphics[width=12cm]{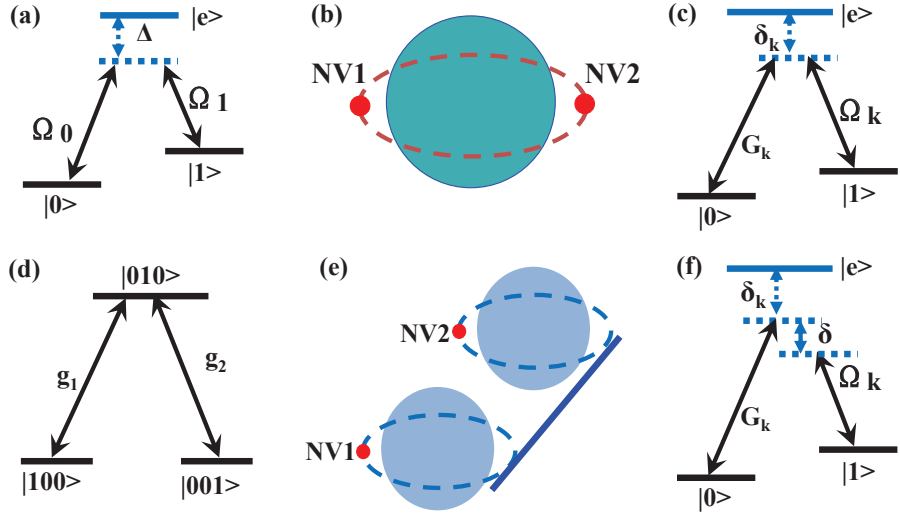}
\caption{The illustration of our proposed implementation: (a) Interaction for single-qubit gates, where three-level configuration of an NV center is driven by two laser fields in a two-photon resonant and a one-photon large detuning of $\Delta$; (b) schematic setup of the nontrivial two-qubit gates in a single cavity case,  with the coupling configuration illustrated in (c), and (e) with optical fiber linked fused-silica microsphere optical cavities,  with the coupling configuration illustrated in (f). Coupling configuration of the $k$th NV center by a laser field with amplitude $\Omega_k$ and coupled to a cavity mode with strength $G_k$  with detunings. (d) In the single excitation subspace, the coupled systems in (b) and (e) are equivalent to a resonant coupled three-level configuration with an effective coupling strength of $g_k$.} \label{setup}
\end{figure*}

\section{Universal single qubit gates}
\subsection{The setup and quantum dynamics}
We first consider the implementation of universal single-qubit gates with holonomies in an all-optical way. The energy level configuration of the NV center solid-state spins is schematically shown in Fig. \ref{setup}(a). The NV center is a defect in a diamond consisting of a substitutional nitrogen atom and an adjacent vacancy, which traps an electron and its electronic ground state has a spin $S=1$. The spin states can be labelled as $|E_{\nu}\rangle\otimes|m_s\rangle$, with $|E_{\nu}\rangle$ and $|m_s\rangle$ being the orbital state, $\nu$ being the angular momentum projection quantum number along the NV center axis, and the spin state with eigenvalue $m_s\hbar$. For our purpose, we choose $|0\rangle=|E_{0}\rangle\otimes|0\rangle$, $|1\rangle=|E_{0}\rangle\otimes|-1\rangle$ and $|e\rangle=(|E_{-1}\rangle\otimes|1\rangle+|E_{-1}\rangle\otimes|-1\rangle)/\sqrt{2}$. Note, the unwanted transition of $|E_{0}\rangle\otimes|1\rangle\leftrightarrow|e\rangle$ can be suppressed using a $\sigma^+$ circular polarization laser, according to the total angular momentum conservation law. In this way, the NV center can be modeled as a three-level $\Lambda$-system \cite{Manson2006}, with the level structure consisting of long lived states $\left\vert i\right\rangle $ ($i=0, 1$) coupled to the spin-orbit excited state $\left\vert e\right\rangle $ by optical driving fields $\Omega_i(t)$, with phase difference $\pi$ \cite{Moller2007}. This selective coupling can be obtained using different polarizations of the driving laser field \cite{Yale2016} but the direct transition between states $\left\vert 0\right\rangle $ and $\left\vert 1\right\rangle $ is electric-dipole forbidden. The initialization, readout, and unitary manipulation of an NV center can be realized all-optically using coherent population trapping and stimulated Raman techniques, the large detuning of which ensures population transfer between the two lower states $\left\vert 0\right\rangle $ and $\left\vert 1\right\rangle $, while the excited state $|e\rangle$ is not populated during the process, thus no loss occurs due to spontaneous emission.

We choose  the parameters of the light-NV-center interaction in a two-photon resonant way as $\Delta_i=\omega_{ei}-\omega_i=\Delta$ with  $\Delta_i$ being single-photon detuning, $\omega_{ei}$ the transitions of $\left\vert e\right\rangle \rightarrow \left\vert i\right\rangle$, and $\omega_i$ as the frequency of driving lasers. In this case, the Hamiltonian of the driving NV center in the basis of $\{|0\rangle, |1\rangle, |e\rangle\}$ can be written as
\begin{eqnarray}\label{h1}
H=\frac{\hbar}{2} \left(\begin{array}{ccc}
0&0&\Omega_0 \\
0&0&-\Omega_1 \\
\Omega_0 &-\Omega_1&2\Delta
\end{array}\right).
\end{eqnarray}
The eigenstates of the this system are
\begin{eqnarray}\label{state}
|d\rangle &=& \cos(\theta/2)|0\rangle + \sin(\theta/2) |1\rangle, \nonumber \\
|+\rangle &=& \sin\varphi |b\rangle + \cos\varphi |e\rangle, \\
|-\rangle &=& \cos\varphi |b\rangle - \sin\varphi |e\rangle, \nonumber
\end{eqnarray}
where $\tan(\theta/2)=\Omega_0/\Omega_1$, $\varphi$ is defined by $\tan{2\varphi}=\Omega/\Delta$ with $\Omega=\sqrt{\Omega_0^2+\Omega_1^2}$, and  $|b\rangle=\sin(\theta/2)|0\rangle - \cos(\theta/2) |1\rangle$  is a bright state.

\subsection{Universal single-qubit gates}

When the pulse shapes of the driving laser fields share the same time-dependence but have different amplitudes, we can keep the parameter $\theta$ as a constant, while the states $|b\rangle$ and $|d\rangle$ are also not time-dependent. In this case, in the basis of $\{ |b\rangle, |e\rangle\, |d\rangle\}$, Eq. (\ref{h1}) can be rewritten as \cite{Vitanov1998}
\begin{eqnarray}\label{heff}
H_1=\frac{\hbar \Omega}{2} ( |b\rangle\langle e|  + |e\rangle\langle b|)
+\hbar \Delta  |e\rangle\langle e|.
\end{eqnarray}
In this dressed-state representation, it is obviously that the dark state $|d\rangle$ decouples from the other states, while the state $|b\rangle$ couples to the excited state $|e\rangle$. Therefore, the evolution operator for this two-photon resonant excitation in the subspace spanned by $\{ |b\rangle, |d\rangle\}$ is
\begin{eqnarray}\label{u12}
U(t)= \left(\begin{array}{ccc}
e^{i\gamma}&0 \\
0&1
\end{array}\right),
\end{eqnarray}
where $\gamma(\tau)=\int_0^\tau (\sqrt{\Delta^2+\Omega^2}-\Delta)/2 dt$ is the effective two-photon pulse area. Therefore, the induced operation is a nonadiabatic holonomic matrix. In the computational space spanned by $\{ |0\rangle, |1\rangle\}$, this can be presented as
\begin{eqnarray}\label{u1}
U(\tau) &=& e^{i\frac{\gamma}{2}} \left(\begin{array}{cc}
\cos \frac{\gamma}{2}-i\sin\frac{\gamma}{2}\cos\theta &-i\sin\frac{\gamma}{2}\sin\theta \\
-i\sin\frac{\gamma}{2}\sin\theta& \cos\frac{\gamma}{2}+i\sin\frac{\gamma}{2}\cos\theta
\end{array}\right) \nonumber \\
&=& e^{i\frac{\gamma}{2}} e^{-i\frac{\gamma}{2} \vec{n} \cdot \vec{\sigma}}
\end{eqnarray}
by applying the Pauli matrix decomposition, where $I$ is the identity matrix, and the unit vector $\vec{n}=\left(\sin\theta, 0, \cos\theta\right)$. From  Eq. (\ref{u1}), one can obtain an arbitrary single-qubit gate. Moreover, the induced gates are of the geometric nature. The dressed states undergo a cyclic evolution  as $|j(\tau)\rangle\langle j(\tau)|=|j(0)\rangle\langle j(0)|$ with $j\in\{b, d\}$. Meanwhile, when $\Omega$ is time-independent, $\langle j(t)|H_1|i(t)\rangle=\langle j|e^{iH_1 t}H_1e^{-iH_1 t}|i\rangle=\langle j|H_1|i\rangle=0$ with $i\in\{b, d\}$, thus the parallel-transport condition is satisfied and the evolution is purely geometric (without dynamical phases).  As the above two conditions are met, the evolution operator $U(\tau)$ is a holonomic gate in our qubit subspace \cite{Sjoqvist2012,Xu2012}. Therefore, the geometric nature of the proposed operation originates from the structure of the Hamiltonian instead of being a result of slow evolution as in the adiabatic case.

For example, controlling the operation time to meet the condition $\gamma(\tau)=\pi$, in the computational space spanned by $\{|0\rangle, |1\rangle \}$ we can realize the nonadiabatic holonomic single-qubit gates as
\begin{equation}\label{u2}
U_{1}(\theta)=\left(\begin{array}{ccc}
\cos{\theta}&\sin{\theta}\\
\sin{\theta}&-\cos{\theta}
\end{array}\right),
\end{equation}
where $\theta$ can be chosen by tuning the amplitude of two lasers, resulting in a set of nonadiabatic one-qubit quantum gates. For example, a Hadamard gate and a NOT gate can be implemented by choosing $U_1(\pi/4)$ and $U_1(\pi/2)$, respectively.

\begin{figure}[tbp]
\centering
\includegraphics[width=8cm]{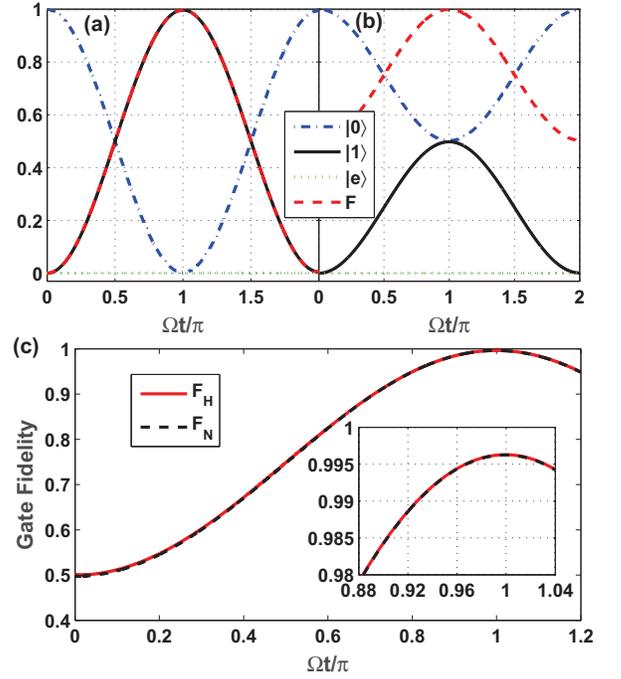}
\caption{ Numerical simulation of the single-qubit operator $U_1(\theta)$ with the initial state set to be $|0\rangle$. State populations and fidelities for the (a) Hadamard gate ($\theta=\pi/4$) and (b) NOT gate ($\theta=\pi/2$). (c) The Hadamard and NOT gate fidelities in initial states are in the form $|\psi\rangle=\cos\Theta|0\rangle+\sin\Theta|1\rangle$ with different $\Theta$.}\label{single}
\end{figure}

\subsection{Gate performance}
Considering the relaxation  and dephasing of the NV center, we simulated the performance of our scheme under realistic conditions using the Lindblad master equation
\begin{eqnarray}\label{master1}
\dot\rho=-i[H_1, \rho]+\frac 1 2 [\gamma_y \mathcal{L}(A^-)+\gamma_x \mathcal{L}(S^-)+\gamma_z \mathcal{L}(S^z)],
\end{eqnarray}
where $H_1$ is the Hamiltonian in the form of Eq. (\ref{h1}), $\rho$ is the density operator, $A^-=|1\rangle\langle 0|$, $S^-=|e\rangle\langle 0| + |e\rangle\langle 1|$, $S^z=|e\rangle\langle e| - |0\rangle\langle 0| - |1\rangle\langle 1|$ and $\mathcal{L}(\mathcal{A})=2\mathcal{A}\rho \mathcal{A}^\dagger-\mathcal{A}^\dagger \mathcal{A} \rho -\rho \mathcal{A}^\dagger \mathcal{A}$
is the Lindblad operator. In our simulation, we used a conservative set of experimental parameters. The Rabi frequency $\Omega=2\pi \times 300$ MHz and $\Delta/\Omega=20$ was used to suppress the excited state population. The qubit relaxation and dephasing rates were estimated to be $\gamma_y \approx 2\pi \times 5$ kHz, $\gamma_x=\gamma_z \approx 2\pi \times 1.5$ MHz \cite{Yale2013,Yale2016}.   Assuming that the qubit was initially prepared in the $|0\rangle$ state, the time-dependent state populations and state fidelities of the Hadamard and NOT gates are depicted  in Fig. \ref{single}(a) and \ref{single}(b), where the obtained state fidelities are $99.92\%$ and $99.65\%$, respectively. Therefore, these high fidelities also verify the validity of adiabatic elimination of the excited state. Furthermore, we investigated the gate fidelity of Hadamard and NOT gates by choosing 1001 different $\Theta$, uniformly distributed within the range $[0, \pi/2]$, with the initial state in the form of $|\psi\rangle=\cos\Theta|0\rangle+\sin\Theta|1\rangle$, as shown in Fig.\ref{single}(c). The gate fidelities of Hadamard and Not gates reached $99.63\%$ and $99.62\%$, respectively, under decoherence.

\section{Nontrivial two-qubit gates}
\subsection{Single cavity case}
We proceeded to implement a nontrivial two-qubit gate where coupling between two NV centers was needed. We considered the scenario where two NV centers were coupled to a fused-silica microsphere optical cavity \cite{Park2006, Barclay2009} with large detuning. A schematic diagram of the nontrivial two-qubit gates is shown in Fig. \ref{setup}(b). This type of microcavity can be made to have small volumes and high-Q factors $Q=8\times10^{9}$ \cite{Vernooy1998}. The lowest-order whispering-gallery mode of the cavity corresponds to the light traveling around the equator of the microsphere, due to continuous total internal reflection. The NV centers can be attached around the equator of the cavity, and light-matter interaction can be induced via the evanescent field of the cavity mode \cite{Park2006}. Here, the NV centers are fixed and the distance between two NV centers is chosen to be much larger than the wavelength ($\sim 0.6$ $\mu$m) of the whispering-gallery mode, so that each driving laser field, with a Rabi frequency $\Omega_k$ ($k=1, 2$), as shown in Fig. \ref{setup}(c), can interact individually with the $k$th NV center and any direct coupling among the NV centers is negligible. The coupling between the two NV centers is mediated by a cavity in the Raman resonant regime. We here employed two three-level NV centers, the same as in the one-qubit case. By choosing the proper parameters, the cavity mode and driving field with frequency $\omega_c$ and $\omega_{d,k}$ couple to the transition $\left\vert j\right\rangle \rightarrow \left\vert e\right\rangle$ with $\omega_{ej,k}$ ($j=0,1$) of the $k$th NV center with a coupling strength of $G_k$ and $(-1)^k\Omega_k$ with phase difference $\pi$, respectively. The detuning $\delta_k=\omega_{e0,k}-\omega_c =\omega_{e1,k}-\omega_{d,k}$ was the same for every NV center.

In the interaction picture with respective to
\begin{eqnarray}\label{h0}
H_0=\sum _{k=1}^{2} \left( \omega_{e0,k}|e\rangle_k\langle e| +(-1)^k\omega_{10,k}|1\rangle_k\langle 1| + \omega_{c,k} a_k^\dagger a_k \right)\notag\\
\end{eqnarray}
with $\omega_{10,k}=\omega_{e0,k}-\omega_{e1,k}$, under the rotating wave approximation, the interaction Hamiltonian reduces to
\begin{eqnarray}\label{hs}
H_I&=&\sum _{k=1}^{2} \left(G_k a|e\rangle_k\langle 0|+ (-1)^k\Omega_k |e\rangle_k\langle 1|\right)e^{i\delta_k t}+ \text{H.c.},
\end{eqnarray}
where $a^{\dag}(a)$ is the creation (annihilation) operator of the cavity.  When $\delta_k \gg \{G_k, \Omega_k\}$, the effective Raman Hamiltonian can be written as
\begin{equation} \label{h2}
H_2=\sum _{k=1}^{2} g_{k}\left(a\sigma _{k}^{+}+\text{H.c}.\right),
\end{equation}
where the effective cavity-assisted coupling strength $g_{k}= (-1)^{k+1}G_k\Omega_k/\delta_k$ can be conveniently tuned via the amplitude of the corresponding external driven laser field $\Omega_k$.

%\subsection{Nontrivial two-qubit gates}
Thus the Hamiltonian in Eq. (\ref{h2}) establishes a resonant three-level $\Lambda$ system in the single excitation subspace $\{|100\rangle, |010\rangle, |001\rangle\}$ of the coupled Hamiltonian, as shown in Fig. \ref{setup}(d), where $|mnq\rangle\equiv|m\rangle_1|n\rangle_c|q\rangle_2$ with the subscript 1, 2 and $c$ indicate the states belonging to NV centers 1 and 2, and the cavity. The effective Rabi frequency $\lambda=\sqrt{g_1^2 + g_2^2}$ and $\vartheta=2\arctan(g_1/g_2)$ can be tuned by the amplitude of the incident lasers. Therefore, $\exp\left[-i\int^{\tau_{2}}_{0}H_{2}dt\right]$ under the $\pi$ pulse criterion $\lambda\tau_{2}=\pi$ can induce nontrivial holonomic two-qubit gates.  In the space spanned by $\{|00\rangle,|01\rangle,|10\rangle, |11\rangle\}$, these gates read
\begin{equation}\label{u2}
U_{2}(\vartheta)=\left(\begin{array}{cccc}
1&0&0&0\\
0&\cos{\vartheta}&\sin{\vartheta}&0\\
0&\sin{\vartheta}&-\cos{\vartheta}&0\\
0&0&0&-1
\end{array}\right),
\end{equation}
where the minus sign of the $|11\rangle\langle11|$ elements comes from the evolution of the dual two-excitation subspaces of $\{|011\rangle, |101\rangle, |110\rangle \}$.

\begin{figure}[tbp]
\centering
\includegraphics[width=7cm]{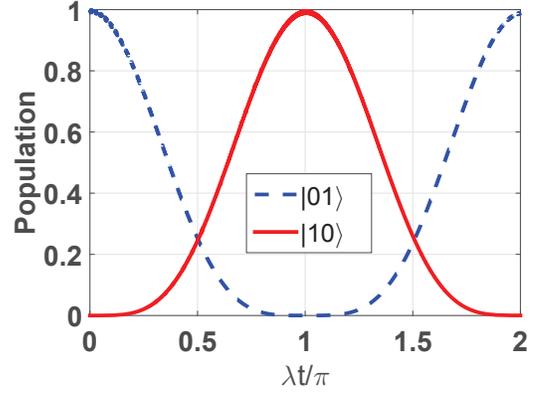}
\caption{State populations and fidelity of the two-qubit gate $U_2(\pi/2)$ with the initial state being $|01\rangle$.}\label{2bit}
\end{figure}

In general, a nontrivial two-qubit holonomic gate can be realized by controlling the $\vartheta$, that is, adjusting the amplitude of one of the two laser fields. For example, a SWAP-like gate can be realized by acting $U_{2}(\pi/2)$ on initial state $|01\rangle$ with fidelity $99.51\%$ as shown in Fig. \ref{2bit}. In the numerical simulation, we set $\Omega_k/\sqrt 2= G_k=2\pi \times 1$ GHz with $\delta_k/\Omega_k=10$ to fulfill the condition $\delta_k \gg \Omega_k$. The qubit relaxation and dephasing rates were the same as in the single-qubit case, and the cavity decay rate was $\kappa = 2\pi c/(\lambda*Q) \approx 2\pi \times 56$ kHz, with $\lambda=670$ nm \cite{Vernooy1998} being the wavelength of the cavity mode.

\subsection{Coupled cavities scenario}
In the above implementation, the distance between the two NV centers was chosen to be much larger than the wavelength of the cavity mode. However, due to the limited size of the cavity, only a small number of NV centers can be attached. To propose a scalable scheme, a coupled cavity scenario can be used, i.e., different cavities are linked by an optical fiber-taper waveguide \cite{fiber}, which can be used to form a two-dimensional lattice configuration for large-scale quantum computation. In this case, quantum information can be transferred from one cavity to another nearby with exceptional high fidelity. Specifically, for the two-coupled cavity case, we considered coupling two cavities via an optical fiber, as shown in Fig. \ref{setup}(e), where each NV center is fixed around a cavity and driven by a laser with Rabi frequency $(-1)^k\Omega_k$ ($k=1, 2$). The detuning was chosen as $\delta_k=\omega_{e0,k}-\omega_{c,k} =\omega_{e1,k}-\omega_{d,k}' -\delta$, as shown in Fig. \ref{setup}(f). In the interaction picture with respect to $H_0$, without coupling between the two cavities, the interaction Hamiltonian can be described by
\begin{eqnarray}
H_I'= \sum_{k=1}^2 G_k a_k|e\rangle_k\langle 0| e^{i\delta_k t}
+ (-1)^k \Omega_k |e\rangle_k\langle 1| e^{i(\delta_k+\delta) t}
+ \text{H.c.}.\notag\\
\end{eqnarray}
When $\delta_k \gg \{\Omega_k, G_k\}$, the effective Raman Hamiltonian can be written as
\begin{equation} \label{htwo}
H_{eff}=\sum _{k=1}^{2} g_{k}' \left(a_k\sigma _{k}^{+} e^{-i\delta t} +\text{H.c}.\right),
\end{equation}
where the effective cavity-assisted coupling strength $g_{k}'= (-1)^{k+1}G_k \Omega_k({1\over \delta_k+\delta} +{1\over \delta_k})/2$ can be conveniently tuned via the amplitude of the corresponding driven laser field $\Omega_k$.

In this case, the two cavity modes are coupled via a common optical fiber mode. Note that, in the short fiber limit, only one resonant mode $b$ of the fiber can interact with the cavity modes. The interaction Hamiltonian reads \cite{fibercouple}
 \begin{eqnarray}
H_c&=&Jb(a_1^\dagger+e^{i\varphi}a_2^\dagger)+\text{H.c}.\notag\\
&=&\sqrt{2}J (c_1^\dagger c_1 - c_2^\dagger c_2),
\end{eqnarray}
where $J$ is the inter-cavity coupling strength, $c=(a_1 - e^{i\varphi} a_2)/\sqrt{2}$,  $c_1=(a_1 + e^{i\varphi} a_2 +\sqrt{2} b)/2$, and  $c_2=(a_1 + e^{i\varphi} a_2-\sqrt{2} b)/2$ are three bosonic normal modes. In the transformed frame with respective to $H_c$ and within the rotating-wave approximation, $H_{eff}$ reads
 \begin{eqnarray} \label{h2'}
H_{2}' =\sum_{k=1}^2 {g_k' \over 2} c_2  \sigma _{k}^{+} +\text{H.c}.,
\end{eqnarray}
where we set $\delta=\sqrt{2}J\gg |g_k'|/2$ and omitted a phase factor of $e^{-i\varphi}$ for $g_2$, which can be compensated for by the phase factor $\Omega_2$. In this case, the two NV centers can be modeled as commonly interacting to a bosonic normal mode, which is in the same form as that of the single cavity case in Eq. (\ref{h2}). For example, $\delta=\Omega_k/2$ leads to $|g_k|\approx G_k/10.24$, which results in $\delta\approx29\times|g_k|/2$, and thus fulfils the requirement of $\delta \gg |g_k'|/2$. Meanwhile, $J=\delta/\sqrt{2}=G_k/2$ is well within realistic experimental situations \cite{fibercouple}.

\section{Discussion}
The proposed NHQC was obtained by dynamical evolution, where only the initial and final states were within the computational subspace. It is different from the usual adiabatic ones, where the instantaneous states are always stay in the computational subspace and the nonabelian holonomies come from suppressing transitions out of the subspace.

As the position and spatial orientation of the NV centers influence their coupling to the cavity, we set their coupling strength to be different without a loss of generality. We emphasize that our scheme works no matter $G_k$ are the same or not, as the effective coupling strength can be tuned by the amplitude of the driven laser $\Omega_k$. Similarly, the transition frequencies of the NV centers can also be different, leading to a difference in the detuning of $\delta_k$ and $\delta$, which can be reset by the frequency of the driven laser. Moreover, the influence of different $\delta_k$ to $g_k$ can also be compensated for by $\Omega_k$.

\section{Conclusion}
In summary, we proposed an NHQC scheme by manipulating NV center electron spins in an all-optical way. By controlling the detuning, amplitude, and phase differences of the driving lasers, we realized arbitrary single-qubit gates as well as nontrivial two-qubit gates with cavity-assisted interaction. The exceptional spin properties of the NV centers and the all-optical manipulation make our scheme a promising candidate for the experimental implementation of high-fidelity NHQC. Therefore, our scheme opens up the possibility of realizing NHQC on solid-state spins characterized by long coherence times and all-optical controls.

\acknowledgments
We thank Dr Jiang Zhang for helpful discussions. This work was supported by the National Fundamental Research Program of China (Grant No. 2013CB921804), the National Key Research and Development Program of China (Grant No. 2016YFA0301803), and the Education Department of Anhui Province (Grant No. KJ2015A299).

\end{document}